# An ANN Based Call Handoff Management Scheme for Mobile Cellular Network

P. P. Bhattacharya[1], Ananya Sarkar[2], IndranilSarkar[3], Subhajit Chatterjee[4]

[1]Department of ECE, Faculty of Engineering and Technology, Mody Institute of Technology & Science (Deemed University), Rajasthan, India
[2]Department of ECE, College of Engineering and Management Kolaghat, West Bengal, India
[3]Department of ECE, Sobhasaria Group of Institutions, Rajasthan, India
[4]Deparment of ECE, Swami Vivekananda Institute of Science and Technology, Barruipur, West Bengal

*Abstract*

*Handoff decisions are usually signal strength based because of simplicity and effectiveness. Apart from the conventional techniques, such as threshold and hysteresis based schemes, recently many artificial intelligent techniques such as Fuzzy Logic, Artificial Neural Network (ANN) etc. are also used for taking handoff decision. In this paper, an Artificial Neural Network based handoff algorithm is proposed and it's performance is studied. We have used ANNhere for taking fast and accurate handoff decision. In our proposed handoff algorithm, Backpropagation Neural Network model is used.The advantages of Backpropagation method are its simplicity and reasonable speed. The algorithm is designed, tested and found to give optimum results.*

*Keywords*

*Handoff; Backpropagation; Artificial Neural Network; Received Signal Strengths; Traffic Intensities.*

## 1. INTRODUCTION

In mobile cellular communication, maintaining continuous communication when the user migrates from one cell to another is done by changing the controlling base station – a process called call Handoff. Handoff involves measurement, decision and execution. In present generation mobile cellular systems, Mobile Station (MS) estimates the signal strengths from each base station and the value of the received signal level is generally affected by three parameters : path loss, shadow fading and small scale fading. Small scale fading has much shorter correlation distance and averaged out over the time scale under consideration [1] and also anti-multipath fading techniques are available now-a-days [2, 3]. Hence, in a system with anti-multipath technique the effect of small scale fading is not normally considered. But in the present work, multipath fading is considered for considering practical scenarios.

In practice, the low speed mobiles may stop after the handoff execution resulting unnecessary handoff. Similarly, the high speed mobiles may move well into the next cell before the handoff execution resulting call termination. Moreover, the signal strength from base station decreases as exp $(-\gamma d)$ where d is the distance of the mobile station from base station and $\gamma$ is the path loss exponent. In uniform propagation environment, $\gamma$ can be taken as constant. But in real environment $\gamma$ may have different values at different places varying from 2 to 6. So, an algorithm based on path loss exponent and user velocity is essential.

DOI : 10.5121/ijwmn.2013.5610        125

As discussed above, handoff characteristics are user velocity dependent. The effect of mobile velocity on handoff performance has been studied by many workers [4,5,6]. Performance metrics such as probability of handoff, average number of handoff, call blocking probability and call completion probability change significantly as user velocity changes. The traffic density in an average urban area generally follows normal distribution [7]. In our country, the average speed in four metro cities e.g, Delhi, Mumbai, Chennai and Kolkata were found to be 30 Km/hr, 25 Km/hr, 25 Km/hr and 22 Km/hr respectively [8]. Due to the sensitivity of handoff performance to path loss exponent, as discussed in the previous section, a variable hysteresis scheme is already proposed [9] where the hysteresis margin is determined as a function of path loss exponent.

In our work, signal strength from the serving and target base stations and traffic intensities of the serving and target base stations are considered. A three layer ANN model [10] is chosen in the design. Signal strengths from the serving and target base stations are estimated using least square estimation method incorporating Rayleigh fading [11]. A Threshold and hysteresis margin based scheme is chosen where handoff decision is taken only when the signal strength from the current base falls below some threshold value and also the signal strength difference between the current and the serving base station is higher than the hysteresis margin so as to avoid ping – pong effect. In the proposed handoff scheme different signal strengths and traffic intensities are considered to find out the position of handoff. Simulation is carried out using C++ language.

## 2. BACK PROPAGATION NEURAL NETWORK

An ANN which is an information-processing paradigm is configured for a particular application through a learning process. In our proposed algorithm, Backpropagation Neural Network is used which is an iterative method that starts with the last layer and moves backward through the layers until the first layer is reached. In this method, the outputs and the errors in outputs are calculated and the weights on the output units are altered. Then the errors in the hidden nodes are calculated and the weights in the hidden nodes are altered. The Backpropagation algorithm changes the weights to minimize the errors. The Backpropagation (**BP**) structure shown in Fig.1 consists of three groups, or layers, of units: a layer of "**input**" units is connected to a layer of "**hidden**" units, which is connected to a layer of "**output**" units. The activity of the input units represents the raw information that is fed into the network. The activity of each hidden unit is determined by both the activities of the input units and the connectivity weights between the input and the hidden units. The behavior of the output units depends on the activity of the hidden units and the weights between the hidden and output units.

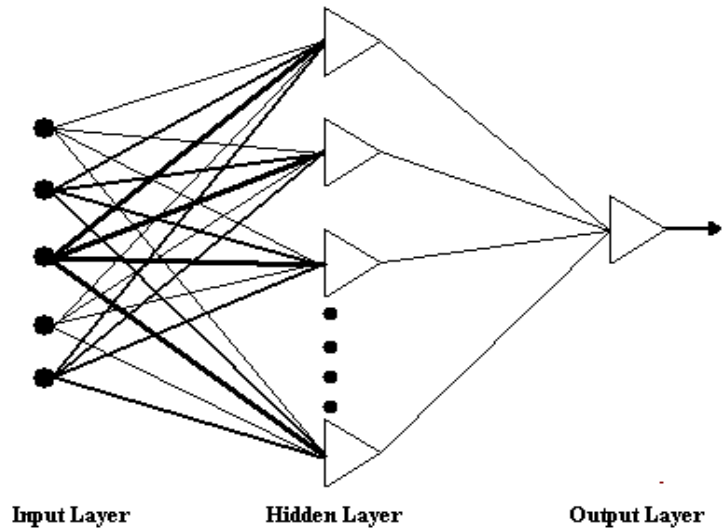

Fig. 1 Backpropagation Neural Network

The number of nodes used in the hidden layer is 20. This number is found after training the network and the errors were found to converge using the value [12]. The output node is a linearly weighted sum of the hidden unit outputs. The outputs decide whether the system needs a handoff or not. If output is -1, no handoff decision will be taken. If output is +1, then handoff decision will be taken. This simple type of network is interesting because the hidden units are free to construct their own representations of the input. The weights between the input and hidden units determine when each hidden unit is active, and so by modifying these weights, a hidden unit can choose what it represents. The advantages of back propagation method are its simplicity and reasonable speed.

Selection of a good activation function is very important because it should be symmetric, and the neural network should be trained to a value that is lower than the limits of the function. One good selection for the activation function is the hyperbolic tangent, or **F(y) = tanh(y)**, because it is completely symmetric, as shown in Fig 2.

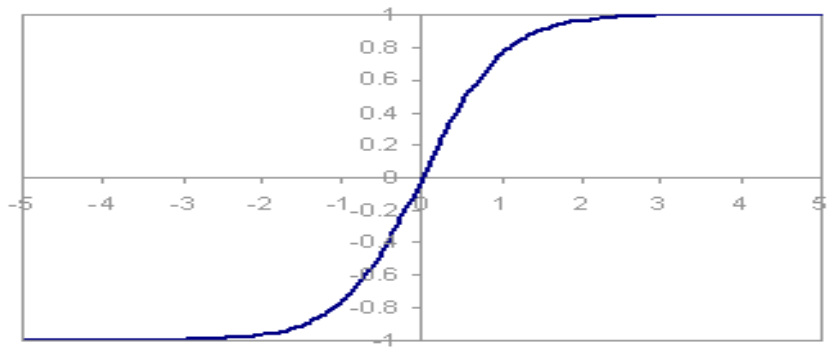

Fig. 2 Hyperbolic tangent function

Another reason for chosing it is that it's easy to obtain its derivative and also the value of derivative can be expressed in terms of the output value (i.e., as opposed to the input value). In our work, this hyperbolic tangent function is chosen.

## 3. PROPOSED HANDOFF ALGORITHM

Two base stations are considered in our paper and the cell radius is assumed to be 500 meter. Fig 3 is the flow chart illustrating the proposed handoff algorithm. Signal strengths of the serving and target base stations are monitored. When the received signal strength from the serving BS less than the threshold value and the received signal strength (**RSS**) from the serving BS is lower than the target BS by hysteresis margin, then a handoff is done to continue the call in progress. Otherwise no Handoff decision will be taken. Then Artificial Neural Network is used to take the handoff decision depending on both **RSS**s and **TI**s [1] of the serving and target BSs. If output of the neural network is +1 then handoff decision should be taken, where as for –1 no handoff will be taken. The threshold value and the hysteresis margin are chosen to be -85 dBm and -5 dBm respectively.

The inputs to the neural network are listed below -

1. $x_1$ is the signal strength of mobile received from the serving BS.
2. $x_2$ is the signal strength of mobile received from the target BS.
3. $x_3$ is the traffic intensity (TI) of the serving BS.
4. $x_4$ is the TI of the target BS.
5. $x_5$ is the bias.

The received signal strength (**RSS**) is considered as:

**Low (L)**: $RSS \leq -85$ dBm and **High (H)**: $RSS > -85$ dBm.

The Traffic Intensity is considered as follow:

**Low (L)**: *TI < 0.66* Erlangs/Channel,
**Medium (M)**: $0.66 \geq TI \geq 0.76$ Erlangs/ Channel,
**High (H)**: *TI >0.76* Erlangs/Channel,

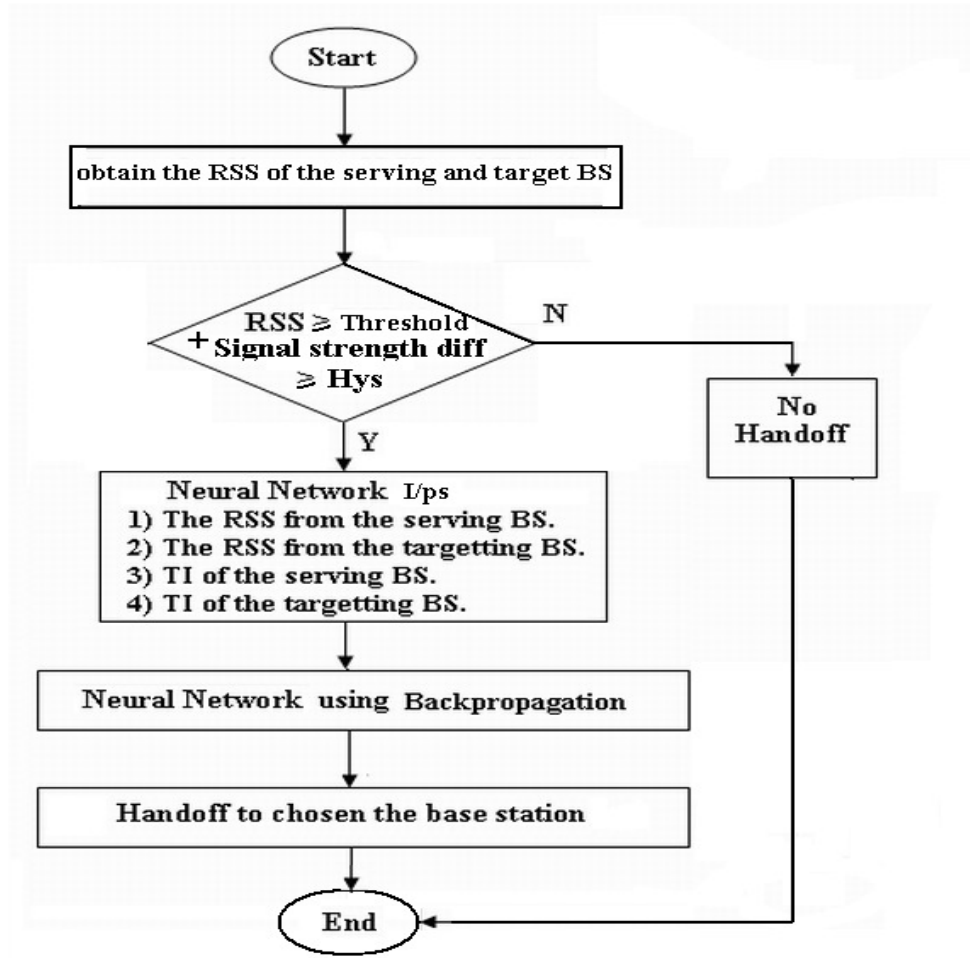

Fig.3 Neural Network based Handoff Algorithm

In this paper four different cases are considered as mentioned in Table 1, such as:

1) Both the **RSS**s from the serving and target **BS**s are low.
2) The **RSS** from the serving **BS** is low while the **RSS** of the target **BS** is high.
3) Both the **RSS**s from the serving and target **BS**s are high.
4) The **RSS** from the serving **BS** is high while the **RSS** from the target **BS** is low.

In each case handoff decisions (**HO**: handoff or **NOHO**: no-handoff) will be taken depending on the different levels of traffic intensities. Let us consider that **RSS** from the serving cell is low and the **RSS** of target cell is high and their traffic intensities are low and high respectively. It is observed that neural network decides not to initiate handoff, as it is desired.

| RSS from the target BS : LOW | RSS from the serving BS : LOW | | | | RSS from the target BS : HIGH | RSS from the serving BS : HIGH | | | |
|---|---|---|---|---|---|---|---|---|---|
| | TI : S / TI : T | LOW | MEDIUM | HIGH | | TI : S / TI : T | LOW | MEDIUM | HIGH |
| | LOW | NOHO | HO | HO | | LOW | NOHO | HO | HO |
| | MEDIUM | NOHO | NOHO | HO | | MEDIUM | NOHO | NOHO | HO |
| | HIGH | NOHO | NOHO | NOHO | | HIGH | NOHO | NOHO | NOHO |
| RSS from the target BS : HIGH | RSS from the serving BS : LOW | | | | RSS from the target BS : LOW | RSS from the serving BS : HIGH | | | |
| | TI : S / TI : T | LOW | MEDIUM | HIGH | | TI : S / TI : T | LOW | MEDIUM | HIGH |
| | LOW | HO | HO | HO | | LOW | NOHO | NOHO | HO |
| | MEDIUM | HO | HO | HO | | MEDIUM | NOHO | NOHO | HO |
| | HIGH | NOHO | NOHO | HO | | HIGH | NOHO | NOHO | NOHO |

Table 1. Handoff decision

## 4. RESULTS AND DISCUSSIONS

The estimated signal strengths [5] from serving and target BS are shown in Fig.4. It is observed that received signal strengthsfluctuatein a random manner in a Rayleigh fading environment. For different values of hysteresis margin and threshold value, the possibilities of handoff against distance from serving base station are calculated for different traffic intensities (Fig.5, 6, 7). It is observed from the Fig.5,that for L/L or M/M or H/H or L/M **(Low :L, Medium :M, High :H)** traffic intensity combinations the distance at which handoff decision is taken remain same. Again for H/L or H/M or M/L traffic intensity combinations the distance at which handoff decision is taken remain same as shown in Fig.6. While for L/H or M/H traffic intensity combinations no handoff will be initiated as shown in Fig.7. The results show quick response and minimum fluctuations in handoff decision. Moreover, average numbers of handoffs versus different hysteresis margins and threshold values are calculated for different sets of traffic intensities (Fig.8.). It is observed from the Fig.8.(a), that for L/L or M/M or H/H or L/M traffic intensity combinations the average numbers of handoffs is same for different hysteresis margins and different threshold values. Again for H/L or H/M or M/L traffic intensity combinations the average numbers of handoffs is same for different hysteresis margins and different threshold values as shown in Fig.8.(b). While for L/H or M/H traffic intensity combinations no handoff will be initiated as shown in Fig.8(c). Thus the algorithm works well under all possible situations.

Hysteresis = -5 dBm and Threshold = - 85 dBm
(Low: L, Medium: M, High: H)

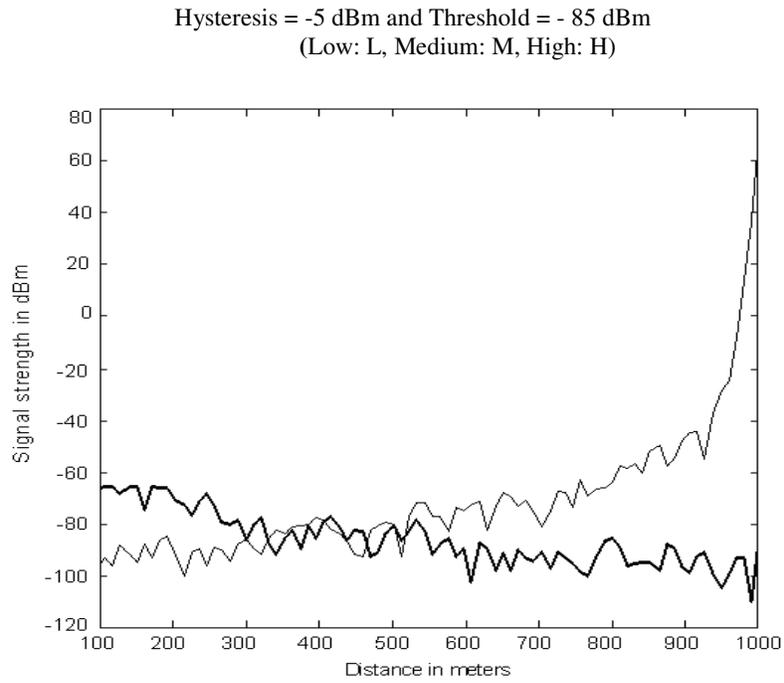

Fig.4 Received signal strengths ofserving and target base stations

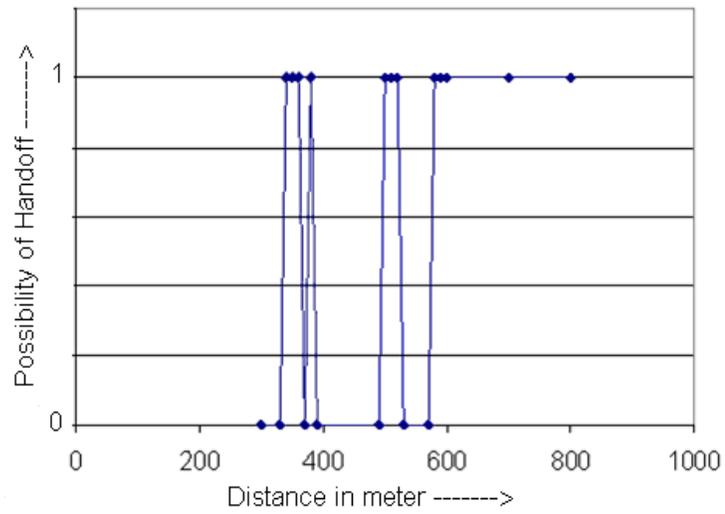

Fig. 5 For L/L or M/M or H/H or L/M   Traffic Intensities

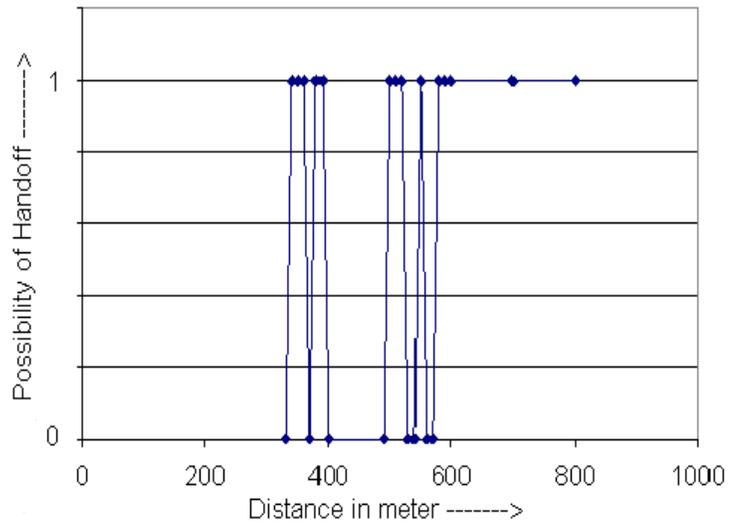

Fig. 6 For H/L or H/M or M/L Traffic Intensities

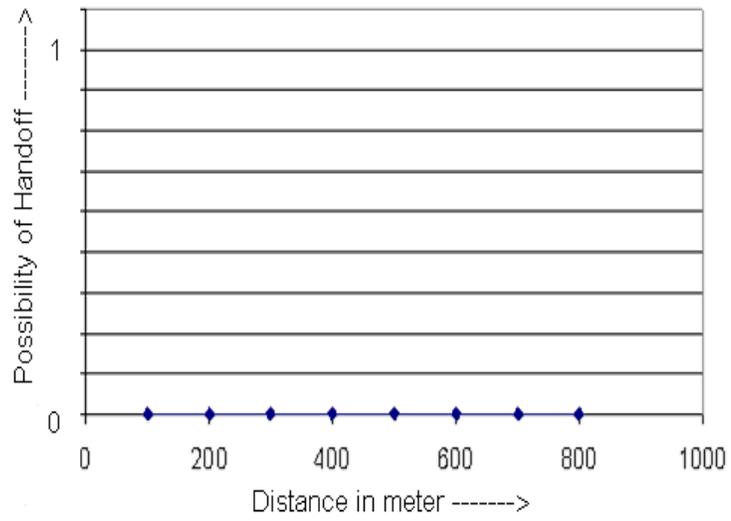

Fig. 7 For L/H or M/H Traffic Intensities

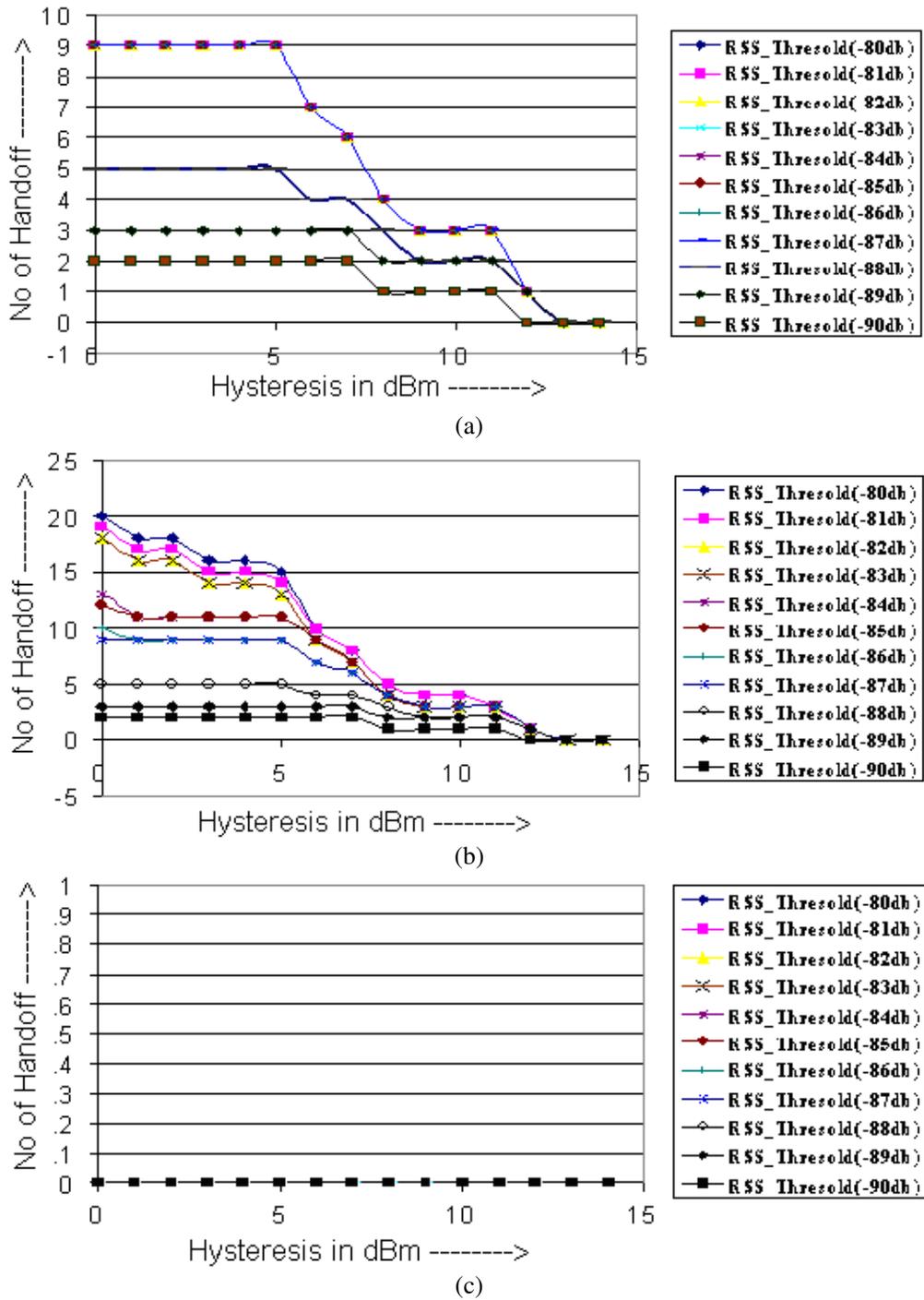

Fig. 8 No of Handoff vs. different Hysteresis margin for different Threshold value (a) For L/L or M/M or H/H or L/M Traffic Intensities (b) For H/L or H/M Traffic Intensities (c) For L/H or M/H Traffic Intensities

## 5. CONCLUSIONS

A handoff algorithm using Artificial Neural Network is designed and the performance of the algorithm is studied in this paper. It is observed from the results that the handoff decisions are taken in appropriate positions and the numberof fluctuation are also low. Average number of handoffis also low which minimizes base station and mobile switching centre processor loading. The designed algorithm can be easily embedded and applied to practical mobile cellular networks.

## REFERENCES

[1]   N. Benvenuto and F. Santucci, "A least square path loss estimation approach to handoff algorithms", IEEE Transaction on Vehicular Technology, vol. 48,  pp 437-447, March 1999.

[2]   Turin, K. L, "Introduction to Spread-Spectrum Antimultipath Techniques and their Application to Urban Digital Radio", Proceedings of the IEEE, vol. 68, no. 3, March 1980, pp. 328 – 353.

[3]   C. Braun, M. Nilson and R. D. Murch, "Measurement of the Interference Rejection Capability of Smart Antennas on Mobile Telephones", IEEE vehicular technology Conference, 1999.

[4]   E. Del. Re, R. Fantacci and G. Giambene, "Handoff and Dynamic Channel Allocation Techniques in Mobile Cellular Networks", IEEE Transactions on Vehicular Technology, Vol 44, No 2, May 1995, pp 229 – 237.

[5]   P. P. Bhattacharya, P. K. Banerjee, "Characterization of Velocity Dependent Mobile Call Handoff", Proc. International Conference on Communication, Devices and Intelligent Systems (CODIS 2004), India, pp. 13-15, 2004.

[6]   B. Venkateswara Rao and Viswanath Sinha, "Study of Channel Assignment Strategies for Handoff and Initial Access in Mobile Communication Networks", IETE Journal of Research, vol 48, No 1, Jan – Feb 2002, pp 69 – 76.

[7]   Traffic Engineering Handbook, Institutes of Transportation Engineering, NOTBO IOA publication, Washington, USA, 1999.

[8]   Sankar Das Mukhopadhyay, "A Comprehensive study of traffic management for the improvement of urban transportation in the districts of West Bengal", Ph.D thesis, Dept. of Business Management, Calcutta University, 2004.

[9]   P. P. Bhattacharya, P. K. Banerjee, "A New Velocity Dependent Variable Hysteresis Margin Based Call Handover Scheme", Indian Journal of radio and Space Physics, Vol. 35, No. 5, pp 368 – 371, October 2006.

[10]  Suleesathira, Raungrong and Kunarak , Sunisa, Non-member,.(2005), "Neural Network Handoff Algorithm in a Joint   Terrestrial-HAPS Cellular System", 164 ECTI transactions on electrical engg., electronics,and communications. VOL.3, NO.2 August 2005

[11]  P. P. Bhattacharya, P. K. Banerjee, "Fuzzy Logic Based Handover Initiation Technique For Mobile Communication In Rayleigh Fading environment", proc. IEEE India annual conference (INDICON 2004), Kharagpur, India, 2004.

[12]  P. P. Bhattacharya, "Application of Artificial Neural Network in Cellular Handoff Management", Proc.  International Conference in Computer Intelligence and Multimedia Applications (ICCIMA 07), Sivakasi, India, 2007

**Authors**

Dr. Partha Pratim Bhattacharya was born in India on January 3, 1971. He has 17 years of experience in teaching and research. He served many reputed educational Institutes in India in various positions. At present he is working as Professor in Department of Electronics and Communication Engineering in the Faculty of Engineering and Technology, Mody Institute of Technology and Science (Deemed University), Rajasthan, India. He worked on Microwave devices and systems and mobile cellular 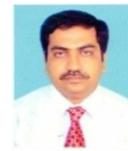 communication systems. He has published more than 100 papers in refereed journals and conferences. His present research interest includes mobile cellular communication, wireless sensor network and cognitive radio.

Dr. Bhattacharya is a member of The Institution of Electronics and Telecommunication Engineers, India and The Institution of Engineers, India. He is the recipient of Young Scientist Award from International Union of Radio Science in 2005. He is working as the editorial board member and reviewer in many reputed journals.

Ananya Sarkar was born in India on April 17, 1978. She received her B. Tech degree in Electronics and Communication Engineering from North Bengal University, India in 2003 and obtained M. Tech degree in VLSI design and Microelectronics Engineering from Jadavpur University, India 2008. She has ten years of experience in teaching and research. At present she is working as an Assistant Professor in the Department of Electronics & Communication Engineering at college of Engineering & Management, Kolaghat, W.B, 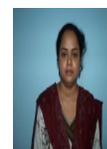 under West Bengal University of Technology (WBUT). Her present research interest   is on Wireless Communication.

Mr. Indranil Sarkar is working as Assistant Professor in Department of Electronics and Communication Engineering in Sobhasaria Group of Institutions (Accredited by NBA & IAO), Rajasthan, India. He has more than 10 years of experience in industry andteaching. He served many reputed industries like WIPRO LTD and educational Institutes in India in various positions starting from Junior Executives to Quality Control Executives  and 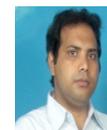 Lecturer to Assistant Professor. He is an associate member of The Institution of Electronics and Telecommunication Engineers, India.  He worked on Microstrip antenna using defected ground structures and mobile cellular communication systems. His field of interest includes microstrip antenna using defected ground structures, mobile communication and cognitive radio.

Subhajit Chatterjee was born in India. He has more than 12 years experience in teaching starting and 3 years in industry . He has served some  reputed educational institutes in different positions from Lecturer to Teacher In Charge of department. Presently he is working as Assistant Professor, Department of Electronics & Communication Engineering, Swami Vivekananda Institute of Science & Technology, West Bengal. He has number of publications and his area of interests  are Analog & Digital Communications, Solid State 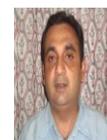 Devices, Telecommunication, Microprocessors & Microcontrollers. He has co authored two books on Microprocessors & Microcontrollers and Telecommunication & Switching. He has reviewed chapters of books published by reputed publishers.Presently he is doing his research on Spectrum Access in Cognitive Radio.